\documentclass[aps,preprint]{revtex4}
\usepackage{graphicx}

\begin{document}

\title{Biology helps to construct weighted scale free networks}

\author{A. Ramezanpour}
\email{ramzanpour@mehr.sharif.edu}
 \affiliation{Department of Physics, Sharif University of
Technology, P.O.Box 11365-9161, Tehran, Iran}
 \affiliation{Institute for Advanced Studies in Basic Sciences, Zanjan 45195-159, Iran}

\date{\today}

\begin{abstract}
In this work we study a simple evolutionary model of bipartite
networks which its evolution is based on the duplication of nodes.
Using analytical results along with numerical simulation of the
model, we show that the above evolutionary model results in
weighted scale free networks. Indeed we find that in the one mode
picture we have weighted networks with scale free distributions
for interesting quantities like the weights, the degrees and the
weighted degrees of the nodes and the weights of the edges.
\end{abstract}

\maketitle

\section{Introduction}\label{1}
Most of interacting systems can be regarded as complex networks
\cite{ab,d,n1}. Certainly, seeking the structural and universal
properties of these networks is a main goal in studying the
behavior of these systems \cite{ws,ba,asbs,ajb,n2}. Among these
one can refer to the small-world phenomenon \cite{ws} and the
scale free behavior of degree distribution \cite{ab}, where degree
denotes the number of nearest neighbors of a node. Clearly finding
the basic ingredients to produce such behaviors helps us in a
better understanding of real networks. For example it is now clear
that a simple evolution of networks in which new nodes prefer to
be connected to higher degree nodes, could give rise to scale free
networks with a power law degree distribution ($P(k)\sim
k^{-\gamma}$) \cite{ba}. The above process seems a natural rule in
the evolution of most of the real networks and one can even
measure the tendency of new nodes to have a preferential
attachment \cite{jnb}.\\
An interesting feature of real networks is that they are complex
weighted networks \cite{bbpv,bbv}. For example we can associate a
weight to each node of a network which might represent the size or
power of that node to create connections with the other nodes. In
a protein complex network this weight is the number of proteins
attributed to a protein complex \cite{g,mrk}. We could also assign
a weight to each edge of a network which might be a measure of
interaction between the end point nodes of the edges in the
network. In the example of protein complex network this weight
shows the number of proteins that two protein complexes have in
common. The weight of an edge in this case would be an appropriate
measure to quantify the functional correlation of two protein
complexes connected by that edge. In the same way one could
consider social collaboration networks, e.g.  scientific
coauthorship networks \cite{n3,bjnrsv,rdp}, as weighted networks.
In this situation the weight of a node, which represents an
author, gives the number of articles written by that author and
the weight of an edge between two nodes is the number of articles
that the corresponding authors have coauthored together. It is
reasonable to think that two authors with a larger number of
articles in common, would have a larger communication and so
would be closer to each other than to the other authors. As
another example one may take the social network of communities or
groups in a society. Each group has a weight which represents the
number of its members and the weight of an edge connecting two
groups gives the number of shared members. Certainly two groups
with a larger number of members in common have a larger
interaction with each other and so a higher probability to
transmit any kind of information between
one another.\\
As the above paragraph reveals, there are a large number of
weighted networks which can be exhibited as a one mode picture of
a bipartite network \cite{nsw,rdp}. For example in the case of
the protein complex network we could make a bipartite network of
proteins and protein complexes. An edge in this bipartite network
only connects a protein (a node of type I) to a protein complex
(a node of type II) and means that this protein is a member of the
associated protein complex. In the same way one can construct the
bipartite network of a scientific coauthorship network where nodes
of type I and II represent the articles and the authors
respectively.\\
The presence of power law distributions with exponents around $2$
is a main characteristic of the above studied networks
\cite{bbpv,mrk,n3,nsw,rdp}. Here the relevant distributions are
the weight, the degree and the weighted degree distribution of the
nodes and the weight distribution of the edges in the one mode
picture. We define the weighted degree of a node as the sum of the
weights of the edges emanating from that node. Thus one can ask if
there is a simple rule for the evolution of bipartite networks
which reproduces the basic features of the above complex weighted
networks?

In this paper we study a simple model for the evolution of
bipartite networks.  To this end we apply a well known rule of
biology in the context of protein evolution, that is {\it
duplication} of proteins,  to the evolution of bipartite networks.
It has been shown that this mechanism can well reproduce the
structural properties of the protein interaction networks
\cite{w,spsk}. In Ref. \cite{mrk} the same procedure has been
applied to simulate the evolution of a protein complex network.
Let us illustrate the duplication mechanism in the example of a
scientific coauthorship network; A new article in this network
could well be assumed as a result of an old article (it has been
duplicated) with some changes probably in the list of the authors
(its connections have undergone mutation). This new article may
also introduce a new author to the list of present authors (it
creates a new node of the other type). Note that an author with a
higher number of articles has a larger probability to produce a
new article. This feature automatically enters the model if we
select randomly an article for duplication. This is an important
property of duplication mechanism in the bipartite networks which
results in the emergence of scale free distributions.\\
In the following we will permit both types of the nodes to have
the chance of duplication. Note that this event is meaningless in
the example of scientific coauthorship network in which the
authors of an article has been fixed at the time of its birth.
However, in other examples such as the social network of groups,
it is a reasonable event where a new group might form as a
duplication of an already present group. Here for the sake of
simplicity we only consider the simple case of pure duplication of
the nodes. Finally we shall take into account the limited age of
the nodes which prevents them from having connections with the new
nodes. This phenomenon has an essential role in networks like the
scientific collaboration networks where a retired person can not
contribute in writing a new article. It turns out that the simple
model introduced in this paper can generate complex weighted
networks with scale free distributions for both the weight of the
nodes and the edges and also for the degree and the weighted
degree of the nodes in the one mode pictures.

The paper is organized as follows. In the next section we give the
model definition in detail. Section \ref{3} is devoted to the
analytic study of the model along with the results of the
numerical simulations. In section \ref{4} we study the effect of
limited ages of the nodes on the behavior of the interesting
quantities by means of numerical simulations. Section \ref{5}
includes the conclusion remarks of the paper.

\section{The model definition}\label{2}
\begin{figure}
\includegraphics[width=8cm]{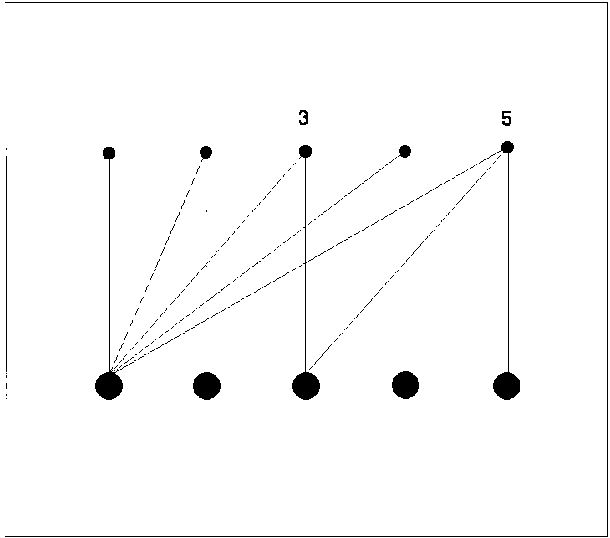}
\caption{A step of the evolution of a bipartite network in which
node $3$ of type I has been duplicated. The result of this
duplication is node $5$.}\label{fig1}
\end{figure}

Consider a bipartite network with $n$ nodes of the first type and
$N$ nodes of the second type. We will indices nodes of type I by
small letters like a,b,c,$\ldots$ and nodes of the other type by
capital letters, that is A,B,C,$\ldots$. In the same way each
quantity will be represented by small or capital letter according
to its relation with the type of nodes. For example by $m_a(t)$
and $m_A(t)$ we mean respectively the weight of a first and a
second type node at time $t$ . Here the weight of a node is the
number of its connections in the bipartite network. To evolve the
network we go through the following rules:\\
i) in each step we choose one type of nodes for duplication. With
probability $\lambda$  a node of the first type an with
probability $1-\lambda$ a node of the second type will be chosen
for duplication.\\
ii) Suppose that we have decided to duplicate a node of the first
type.  Now a node is randomly chosen to produce a copy of itself.
This copy, which is the same type as the duplicated node, will
also have the same weight and even the same set of connections
with the other type of nodes.\\
iii) Finally we allow the new node to create a node of the other
type and to connect to it. These processes have
been shown in Fig. \ref{fig1}.\\
Note that all the above processes occur in one time step that is
from  $t$ to $t+1$ and the same events could happen for a node of
the second type.

\section{Analytic study of the model}\label{3}
\begin{figure}
\includegraphics[width=8cm]{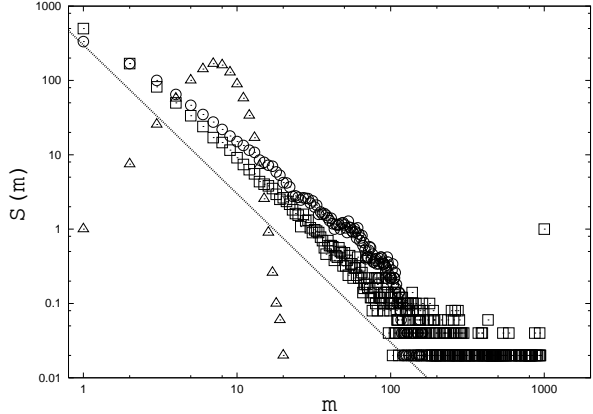}
\caption{Weight distribution of the nodes of type II for some
values of $\lambda$. The parameters are $t=1000$, $\lambda=1$
(squares), $\lambda=0.5$ (circles) and $\lambda=0$ (triangles).
The data are result of averaging over $50$ runs of the evolution
of the model. This number is the same in all the numerical
simulations of the model represented in this paper. The guideline
shows a power law behavior of exponent $2$.}\label{fig2}
\end{figure}

\begin{figure}
\includegraphics[width=8cm]{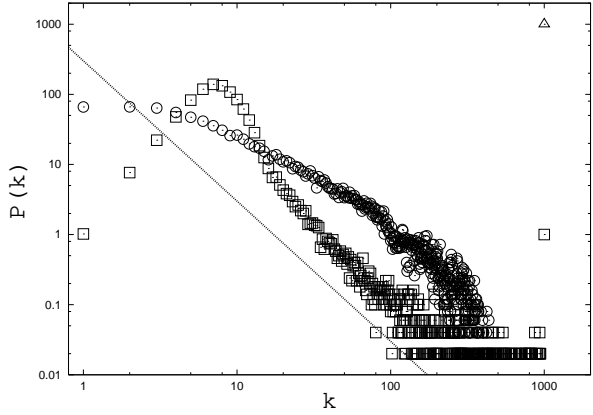}
\caption{Degree distribution of the nodes of type II in the one
mode picture. The parameters are $t=1000$, $\lambda=1$ (squares),
$\lambda=0.5$ (circles) and $\lambda=0$ (triangles). The guideline
shows a power law behavior of exponent $2$.}\label{fig3}
\end{figure}

\begin{figure}
\includegraphics[width=8cm]{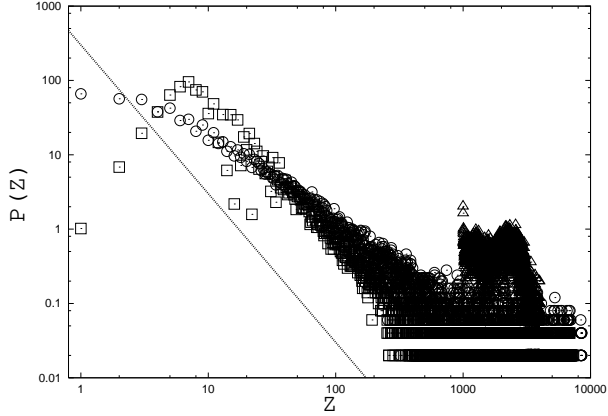}
\caption{ Distribution of weighted degree of the nodes of type II
in the one mode picture. The parameters are $t=1000$, $\lambda=1$
(squares), $\lambda=0.5$ (circles) and $\lambda=0$ (triangles).
The guideline shows a power law behavior of exponent
$2$.}\label{fig4}
\end{figure}

\begin{figure}
\includegraphics[width=8cm]{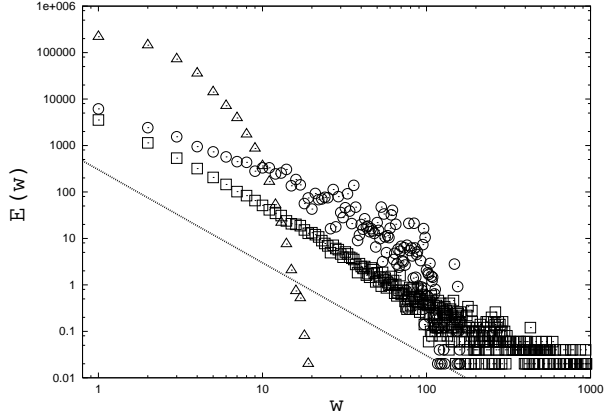}
\caption{Weight distribution of the edges in the one mode picture
of the nodes of type II. The parameters are $t=1000$, $\lambda=1$
(squares), $\lambda=0.5$ (circles) and $\lambda=0$ (triangles).
The guideline shows a power law behavior of exponent $2$.
}\label{fig5}
\end{figure}

Note that the only parameter of the model defined above is
$\lambda$. Moreover due to the symmetry of the evolution if we
compute the behavior of nodes of type II we could get the behavior
of the
other type only by replacing $\lambda$ with $1-\lambda$.\\
As the initial condition we start at time $t=1$ with one node of
type I which has been connected to one node of type II. Thus
according to the deterministic creation of the nodes, the number
of nodes at time $t$ will be given by $n(t)=N(t)=t$.\\
We can write the following equation for the average weight of node
$A$ which entered the network at time $t_A \leq t$:
\begin{equation}\label{M0}
m_A(t+1)=m_A(t)+\lambda m_A(t)/n(t).
\end{equation}
Indeed the second term in this equation is the probability that a
node of type I which is connected to node $A$, be selected for
duplication. If so the weight of node $A$ will increase by one due
to the connection with the new node. Note that a node of type II
has the average weight
\begin{equation}
m_A(t_A)=1+(1-\lambda)\sum_{B=1}^{N(t_A-1)} m_B(t_A-1)/N(t_A-1).
\end{equation}
at the time of its birth. From these equations one can find the
following equation for $\Omega(t):=\sum_A m_A(t)=\sum_a m_a(t)$
\begin{equation}
\Omega(t+1)=\Omega(t)+1+\Omega(t)/t,
\end{equation}
Where $\Omega(t)$ gives the average number of edges in the
bipartite network. Here we have used the fact that $n(t)=N(t)=t$.
In the continuum approximation where we take $t$ as a continuum
variable, the above equation can be rewritten as
\begin{equation}
\frac{d\Omega(t)}{dt}=1+\Omega(t)/t,
\end{equation}
which has the solution
\begin{equation}\label{L0}
\Omega(t)=t(1+\ln t),
\end{equation}
with the initial condition $\Omega(1)=1$. Thus the average weight
of a node in the network will increase with the logarithm of $t$.
In the same way the solution of Eq. \ref{M0} takes the following
form in the continuum approximation
\begin{equation}
m_A(t)=m_A(t_A)(t/t_A)^{\lambda}=\left[1+(1-\lambda)(1+\ln t_A)
\right](t/t_A)^{\lambda}.
\end{equation}
For $\lambda=1$ this equation gives
\begin{equation}\label{m1}
m_A(t)=t/t_A.
\end{equation}
Note that having in hand this behavior we can use the conservation
of probabilities to compute the distribution function of the
weight of the nodes $S(m)$ in the network. Indeed the number of
nodes whose weights are between $m$ and $m+\Delta m$, i.e.
$S(m)\Delta m$, is equal to the number of nodes which have
entered the network between times $t_A$ and $t_A+\Delta t_A$ where
$t_A$ is given by Eq. \ref{m1}. Note also that in each step a new
node has been introduced to the network. Therefore we find that
$S(m)$ behaves like
\begin{equation}
S(m) \sim m^{-2}.
\end{equation}
On the other hand we have the following relation for $m_A(t)$ in
the case of $\lambda=0$
\begin{equation}
m_A(t)=2+\ln t_A,
\end{equation}
which is independent of $t$ and only depends on the birth time of
the node $A$. It is easy to show that in this case
\begin{equation}
S(m) \sim e^{m}.
\end{equation}
Thus the number of nodes increases exponentially with the weight
of node. But the network is finite and we will encounter the
finite size effects for the smaller values of $m$ in this case.
Therefore for $\lambda=1$ we have the exponent $2$ for $S(m)$
while in the case of $\lambda=0$ this exponent will be $\infty$
which reflects the exponential decay of the weight distribution
due to the finite size of the network. These arguments have been
confirmed in Fig.\ref{fig2} which shows the results of numerical
simulations in these cases. Note that when $\lambda=0$ we have
always a node of the second type with weight $t$. This node is
indeed the one we have started with it. We remind that in the
same time the exponent of $s(m)$, the weight distribution of
nodes of type I,
is obtained by interchanging $\lambda$ and $1-\lambda$.\\
Now let us consider the one mode picture of the above bipartite
network constructed by the second type nodes. First we focus on
the evolution of the degree of a node in this picture. Indeed the
number of neighbors of node $A$ increases by one when the
duplicated node is a member of it in the case of first type
duplication. The probability for this to happen is $\lambda
m_A(t)/n(t)$. In the case of second type duplication, the number
of neighbors increases only when the duplicated node is a
neighbor of node $A$ or the node $A$ itself. This probability is
given by $(1-\lambda)(k_A(t)+1)/N(t)$. So for node $A$ with
$t_A\leq t$, we have
\begin{equation}\label{KA0}
k_A(t+1)=k_A(t)+\lambda m_A(t)/n(t)+(1-\lambda)(k_A(t)+1)/N(t).
\end{equation}
In the same way one obtains the average degree of node $A$ at the
time of its birth
\begin{equation}\label{KA}
k_A(t_A)=\lambda
\sum_{b=1}^{n(t_A-1)}m_b(t_A-1)/n(t_A-1)+(1-\lambda)\sum_{B=1}^{N(t_A-1)}k_B(t_A-1)/N(t_A-1).
\end{equation}
But $\sum_{b=1}^{n(t)}m_b(t)=\Omega(t)$ and $\Omega(t)$ is given
by Eq. (\ref{L0}). We also define $L(t):=\sum_{B=1}^{N(t)}k_B(t)$
and use Eqs.(\ref{KA0}) and (\ref{KA}) to write the following
relation for $L(t)$
\begin{equation}
L(t+1)=L(t)+2\lambda
\Omega(t)/n(t)+2(1-\lambda)L(t)/N(t)+(1-\lambda),
\end{equation}
where $L(t)/2$ gives the number of edges in the one mode picture
of the nodes of type II. Solving this equation in the continuum
approximation we find
\begin{equation}
L(t)=(1+\lambda-2\lambda^2)(t^{2(1-\lambda)}-t)/(1-2\lambda)^2-2\lambda
t\ln t/(1-2\lambda).
\end{equation}
For $\lambda=1$ we get $L(t)=2t\ln(t)$ and for $\lambda=0$ this
behavior is replaced by $L(t)=t(t-1)$. Going back to
Eq.(\ref{KA0}) we are now ready to solve it in the continuum
approximation
\begin{equation}
k_A(t)=[1+(1-\lambda)(1+\ln
t_A)](t/t_A)^{\lambda}/(2\lambda-1)-1+C_At^{1-\lambda},
\end{equation}
where $C_A$ is a constant determined by Eq. (\ref{KA}). For
$\lambda=1$ the above relation takes the form
\begin{equation}
k_A(t)=t/t_A+\ln(t_A-1),
\end{equation}
which for $t\rightarrow \infty$ predicts a power law degree
distribution of exponent $2$ for the large values of $k$ , that is
$P(k)\sim k^{-2}$. In Fig.\ref{fig3} we have shown the degree
distribution for some values of $\lambda$. Note that for
$\lambda=0$ we have a fully
connected network in which $k_A(t)=t-1$ for all the nodes.\\
Note also that each edge of the above network has a weight $w$ and
so one can speak of weighted degree of a node $Z(t)$, which gives
the sum of weights of the edges emanating from that node, i.e.
$Z_A(t)=\sum_{B\neq A}w_{AB}(t)$. The average of $Z_A(t)$ for $t_A
\leq t$ is determined by the following considerations; First
consider the case of type I duplication where $m_A(t)/n(t)$ gives
the probability that a member of node $A$ is selected for
duplication. In this case $Z_A(t)$ increases by $m(a|A;t)$ that we
define as the average weight of a node of type I at time $t$ which
is connected to node $A$ in the bipartite network. On the other
hand in the case of the second type duplication, $Z_A(t)$
increases only when the selected node is the node $A$ or one of
its neighbors in the one mode picture. In the latter case $Z_A(t)$
increases by $w(B|A;t)$ which denotes the average weight of an
edge emanating from node $A$ in the one mode picture and in the
former case it increases by $m_A(t)$. Thus we obtain
\begin{equation}\label{Z0}
Z_A(t+1)=Z_A(t)+\lambda
m_A(t)m(t|A)/n(t)+(1-\lambda)\left[m_A(t)/N(t)+k_A(t)w(t|A)/N(t)
\right].
\end{equation}
Similarly when node $A$ enters the network we have
\begin{equation}\label{ZA}
Z_A(t_A)=\lambda
\sum_{a}m_a(t_A-1)/n(t_A-1)+(1-\lambda)\sum_{B}\left[m_B(t_A-1)+k_B(t_A-1)w(t_A-1|B)\right]/N(t_A-1).
\end{equation}
Let us define $Q(t):=\sum_A Z_A(t)$. Then Using Eqs. (\ref{Z0})
and (\ref{ZA}) along with $n(t)=N(t)=t$ we find
\begin{equation}\label{Q}
Q(t+1)=Q(t)+(2-\lambda)Q(t)/t+2\Omega(t)/t,
\end{equation}
where we have used the following relations
\begin{equation}
Z_A(t)=k_A(t)w(t|A)=m_A(t)[m(t|A)-1].
\end{equation}
We can solve Eq. (\ref{Q}) in the continuum approximation and with
the initial condition $Q(1)=0$ to find
\begin{equation}
Q(t)=2(2-\lambda)t(t^{1-\lambda}-1)/(1-\lambda)^2-2t\ln t
/(1-\lambda).
\end{equation}
Now from Eq. (\ref{ZA}) we can write
\begin{equation}\label{ZAA}
Z_A(t_A)=\Omega(t_A-1)/(t_A-1)+(1-\lambda)Q(t_A-1)/(t_A-1).
\end{equation}
Solving Eq.(\ref{Z0}) in the continuum approximation we obtain
\begin{equation}
Z_A(t)=C_At-m_A(t)/(1-\lambda),
\end{equation}
where $C_A$ is again a constant determined by Eq.(\ref{ZAA}). For
$\lambda=1$ it is easy to find that
\begin{equation}
Z_A(t)=\left[\ln t +1 +\ln [(t_A-1)/t_A]\right]t/t_A.
\end{equation}
As it is seen for $t \rightarrow \infty $ we expect a power law
distribution for weighted degrees with exponent $2$, i.e.
$P(Z)\sim Z^{-2}$. On the other extreme that is for $\lambda=0$
we have
\begin{equation}
Z_A(t)=\left[4(t_A-1)-1 -\ln [(t_A-1)/t_A]\right]t/t_A-2-\ln t_A.
\end{equation}
Thus for large times $Z_A(t)$ is nearly independent of $t_A$ and
we find a delta like distribution for this quantity. The reader
can check these statements in Fig.\ref{fig4}.\\
Finally let us study the behavior of weight of the edges in the
one mode picture. The average weight of the edge between two nodes
$A$ and $B$ with $t_A < t_B \leq t$, increases by one only when
the duplicated node is of type I and moreover is connected to
both the nodes. This probability is given by $\lambda
w_{AB}(t)/n(t)$ thus we find
\begin{equation}\label{w0}
w_{AB}(t+1)=w_{AB}(t)+\lambda w_{AB}(t)/n(t).
\end{equation}
Moreover, when node $B$ enters the network at time $t_B$, the
average of its connection weight with a previously present node
is given by
\begin{equation}
w_{AB}(t_B)=\lambda m_A(t_B-1)/n(t_B-1)+(1-\lambda)
\left[m_A(t_B-1)+\sum_{C\neq A}w_{AC}(t_B-1)\right]/N(t_B-1).
\end{equation}
Using the fact that $n(t)=N(t)=t$ and $Z_A(t)=\sum_{C\neq
A}w_{AC}(t)$ we find
\begin{equation}
w_{AB}(t_B)=\left[m_A(t_B-1)+(1-\lambda)Z_A(t_B-1)\right]/(t_B-1).
\end{equation}
Thus taking advantage of the continuum approximation to solve Eq.
(\ref{w0}) we find that the average weight of the edge between
nodes $A$ and $B$ with $t_A<t_B\leq t$ is
\begin{equation}\label{w1}
w_{AB}(t)=w_{AB}(t_B)(t/t_B)^{\lambda}.
\end{equation}
For $\lambda=1$ we have
\begin{equation}
w_{AB}(t)=t/(t_At_B),
\end{equation}
where $t_A$ and $t_B$ can take integer values from $1$ to $t$. Let
us define
\begin{equation}
G(x):=\sum_{t_A=1}^{t-1}\sum_{t_B=t_A+1}^{t}\delta_{x,t_At_B},
\end{equation}
which is the number of edges in the network with $x=t_At_B$. It is
easy to see graphically that $G(x)\simeq x-\sqrt{x}$ for $1<x\leq
t$ and $G(x)\simeq t-\sqrt{x}$ for $t\leq x\leq t^2$. Now we can
use conservation of probabilities
\begin{equation}
\Delta G(x)=(1-1/(2\sqrt x))\Delta x=-E(w) \Delta w.
\end{equation}
to find the behavior of $E(w)$, the distribution of weight of the
edges, for large values of $w$
\begin{equation}
E(w) \sim (t/w^2)(1-\sqrt{w/(4t)}).
\end{equation}
Obviously for large $t$ the exponent of this distribution is $2$.
On the other hand for $\lambda=0$ from Eq. (\ref{w1}) we see that
$w_{AB}(t)= w_{AB}(t_B)$. As before we expect an exponential tail
for the weight distribution of the edges in this case. These
arguments are
confirmed by virtue of the numerical simulations in Fig.\ref{fig5}.\\

\section{Role of limited lifetime}\label{4}
\begin{figure}
\includegraphics[width=8cm]{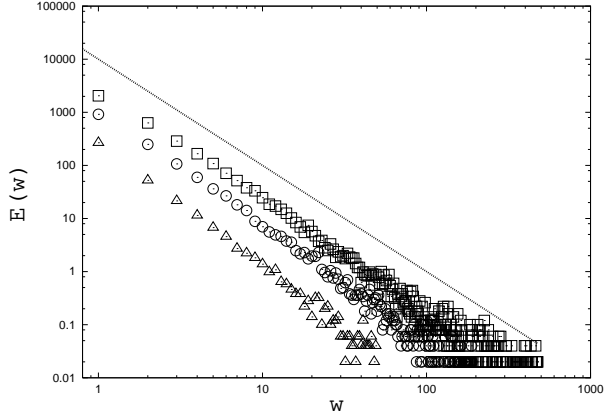}
\caption{Weight distribution of the edges in the one mode picture
of the nodes of kind II when $t^*=\infty$. The parameters are
$t=1000$, $\lambda=1$, $T^*=500$ (squares), $T^*=200$ (circles)
and $T^*=50$(triangles). The guideline shows a power law behavior
of exponent $2$. }\label{fig6}
\end{figure}

\begin{figure}
\includegraphics[width=8cm]{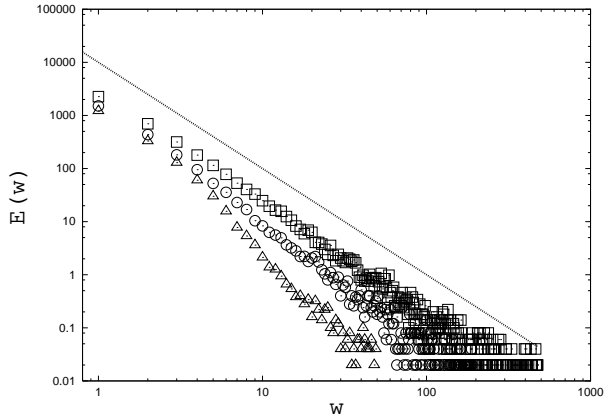}
\caption{Weight distribution of the edges in the one mode picture
of the nodes of kind II when $t^*=T^*$. The parameters are
$t=1000$, $\lambda=1$, $T^*=500$ (squares), $T^*=200$ (circles)
and $T^*=50$(triangles). The guideline shows a power law behavior
of exponent $2$. }\label{fig7}
\end{figure}

In this section we are going to investigate the effect of limited
age of the nodes on the behavior of the distributions studied in
the previous section. To this end we assign a lifetime to each
type of the nodes. That is a node will be active only during its
life which is $t^*$ or $T^*$ according to the type of the node. It
is the only feature that we add to the model studied above. In
this way only the active nodes of each type will have the
opportunity to be selected for duplication. Moreover the new node
can only establish connections with the active nodes of the other
type. Evolving the network in this manner, the number of active
nodes of each type during the evolution will be always less than
or equal to the assigned lifetimes. Nevertheless the total number
of nodes of each type is as before equal to $t$. To see the role
of the limited ages we consider the case of $\lambda=1$ with i)
$t^*=\infty$ and ii) $t^*=T^*$. Since the qualitative behavior of
interesting distributions is the same, we shall only focus on
$E(w)$, the weight distribution of the edges in the one mode
picture of the nodes of type II. Again as the initial condition we
start with a node of type I which has been connected to a node of
type II. In Figs. \ref{fig6} and \ref{fig7} we show the above
distribution for some values of $T^*$. As Fig. \ref{fig6} shows,
by decreasing $T^*$ the general behavior of distribution dose not
change and even its exponent remains nearly constant. Of course
the number of edges with weight zero increases as required by the
conservation of the probability. However in Fig.\ref{fig7} we see
that by decreasing $T^*$ the power law behavior of $E(w)$ slowly
converts to an exponential decay.

\section{Conclusion}\label{5}
In summary we have shown how weighted scale free networks could be
generated by the evolution of bipartite networks which their
evolution is based on a well known rule of biology, that is
duplication of the nodes. We showed that by tuning $\lambda$ which
controls the rate of duplication of the nodes of different types,
one can go from a power law regime to an exponential one where the
tail of the distributions fall off exponentially. In this model
the exponents of interesting distributions are less than or equal
to $2$ and this is close to what seen in the real weighted
networks. We also studied the effect of limited age for the nodes
and showed that a short lifetime may destroy
the power law behavior of distributions.\\
We emphasize that the simple model studied in this paper is a toy
model and far from the evolution of the real networks.
Nevertheless its success in generating scale free distributions
for the important quantities of the weighted networks, indicates
to the essential role of duplication mechanism in the evolution of
complex weighted networks. Certainly one can enrich the above
model, e.g. by introducing mutation to the model , to get a more
realistic evolution.

\acknowledgments

The author is grateful to V. Karimipour for careful reading of the
manuscript and useful suggestions.


\begin{thebibliography}{prsty}

\bibitem{ab}
 R. Albert and A.-L.Barab\'{a}si, Rev. Mod. Phys. \textbf{74},47-97 (2002).

\bibitem{d}
S.N.Dorogovtsev and J.F.F.Mendes, Evolution of Networks : From
Biological Nets to the Internet and WWW, (Oxford University Press,
2003).

\bibitem{n1}
M.E.J.Newman, SIAM Review \textbf{45}, 167-256 (2003).

\bibitem{ws}
 D.J.Watts and S.H.Strogatz , Nature \textbf{ 393},440 (1998).

\bibitem{ba}
 A.-L.Barab\'{a}si and R. Albert, Science \textbf{286}, 509 (1999).
\bibitem{asbs}

L.A.N.Amaral, A. Scala, M.Barth\'{e}l\'{e}my, and H.E.Stanly,
Proc. Natl. Acad. Sci USA \textbf{97},11149(2000).

\bibitem{ajb}
R.Albert, H.Jeong, and A.-L.Barab\'{a}si, Nature \textbf{406},
378(2000)

\bibitem{n2}
M.E.J.Newman, Phys. Rev. Lett. \textbf{89}, 208701 (2002).

\bibitem{jnb}
H. Jeong, Z. Neda and A.-L. Barabasi, cond-mat/0104131.


\bibitem{bbpv}
A. Barrat, M. Barthelemy, R. Pastor-Satorras and A. Vespignani,
Proc. Natl. Acad. Sci. USA \textbf{101}, 3747 (2004)


\bibitem{bbv}
A. Barrat, M. Barthelemy, A. Vespignani, Phys. Rev. Lett.
\textbf{92}, 228701 (2004).


\bibitem{g}A.-C.Gavin et al., Nature \textbf{415}, 141-147(2002).

\bibitem{mrk}
A. Mashaghi, A. Ramezanpour and V. Karimipour, cond-mat/0304207.


\bibitem{n3}
M. E. J. Newman,  Proc. Natl. Acad. Sci. USA \textbf{98}, 404-409
(2001)


\bibitem{bjnrsv}
A.L. Barabasi, H. Jeong, Z. Neda, E. Ravasz, A. Schubert and T.
Vicsek,  Physica A \textbf{311}, (3-4) 590-614 (2002).


\bibitem{rdp}
J. J. Ramasco, S. N. Dorogovtsev and R. Pastor-Satorras,
cond-mat/0403438.

\bibitem{nsw}
M.E.J.Newman, S.H.Strogatz and D.J.Watts, Phys. Rev. E
\textbf{64}, 026118 (2001).

\bibitem{w}A.Wagner,
Mol.Biol.Evol. \textbf{18}(7):1283-1292(2001).

\bibitem{spsk}
R. V. Sole, R. Pastor-Satorras, E. Smith and T. B. Kepler,
Advances in Complex Systems \textbf{5}, 43 (2002).



\end{thebibliography}
\end{document}